\documentclass[11pt]{article}
\usepackage{moriond,epsfig}

\def\be{\begin{equation}}
\def\ee{\end{equation}}
\def\bea{\begin{eqnarray}}
\def\eea{\end{eqnarray}}

\begin{document}
\vspace*{4cm}
\title{TOP QUARK MASS MEASUREMENTS AT THE TEVATRON}

\author{ U. HEINTZ }

\address{Boston University, Department of Physics, 590 Commonwealth Avenue, Boston, MA 02215, USA}

\maketitle\abstracts{
The mass of the top quark has been measured at Fermilab using the Tevatron $p\bar{p}$ collider to a precision of less than 1\%. I discuss the individual measurements that go into this result and its impact on our understanding of the standard model of the electroweak interactions.}

\begin{figure}[h]
\vspace{-0.2in}
\begin{minipage}{0.45\linewidth}
\centering
\includegraphics[width=2.9in]{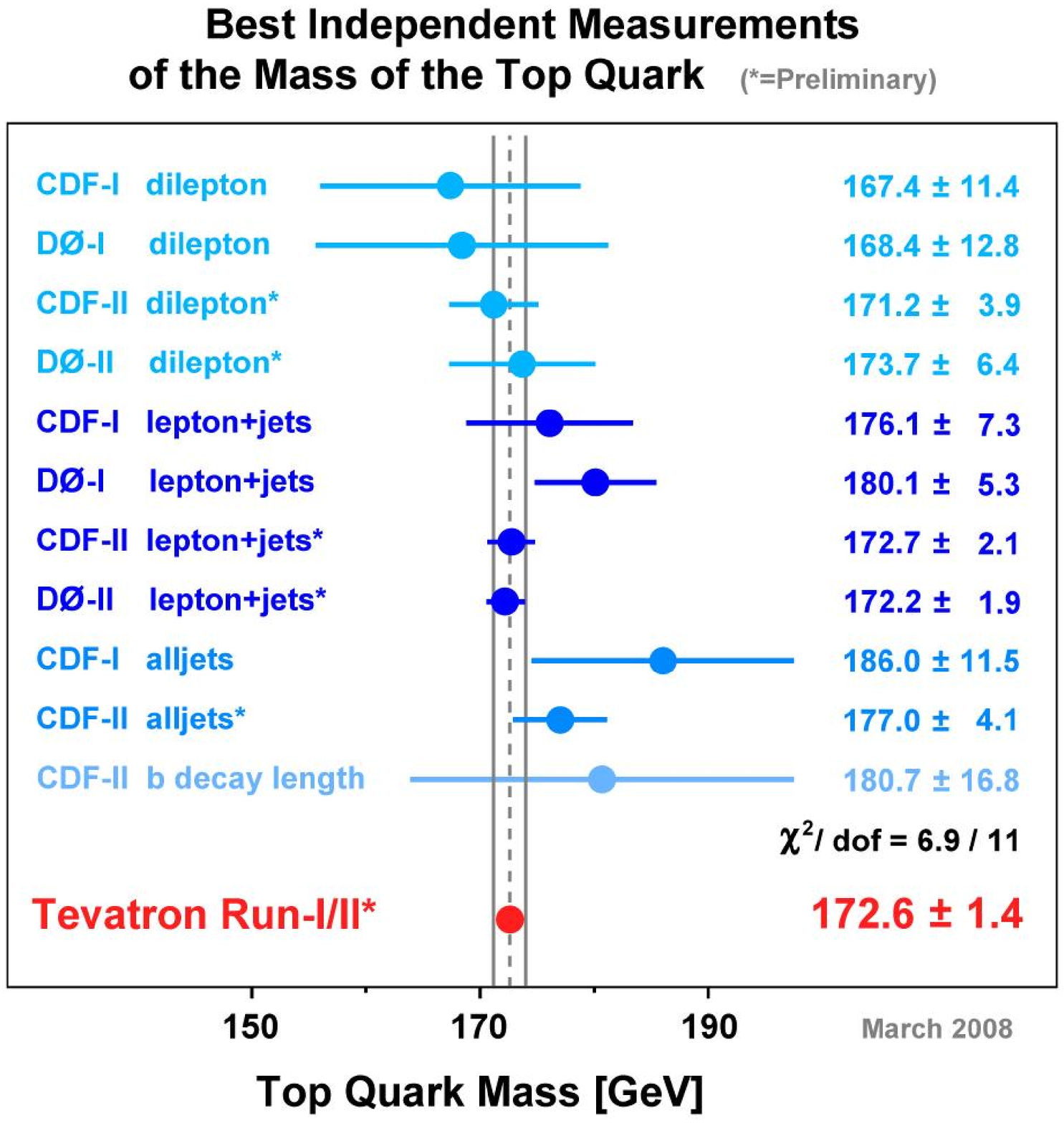}
\caption{Measurements of the top quark mass.}
\label{fig:mtop} 
\end{minipage}%
\hfill
\begin{minipage}{0.45\linewidth}
\centering
\includegraphics[width=3.1in]{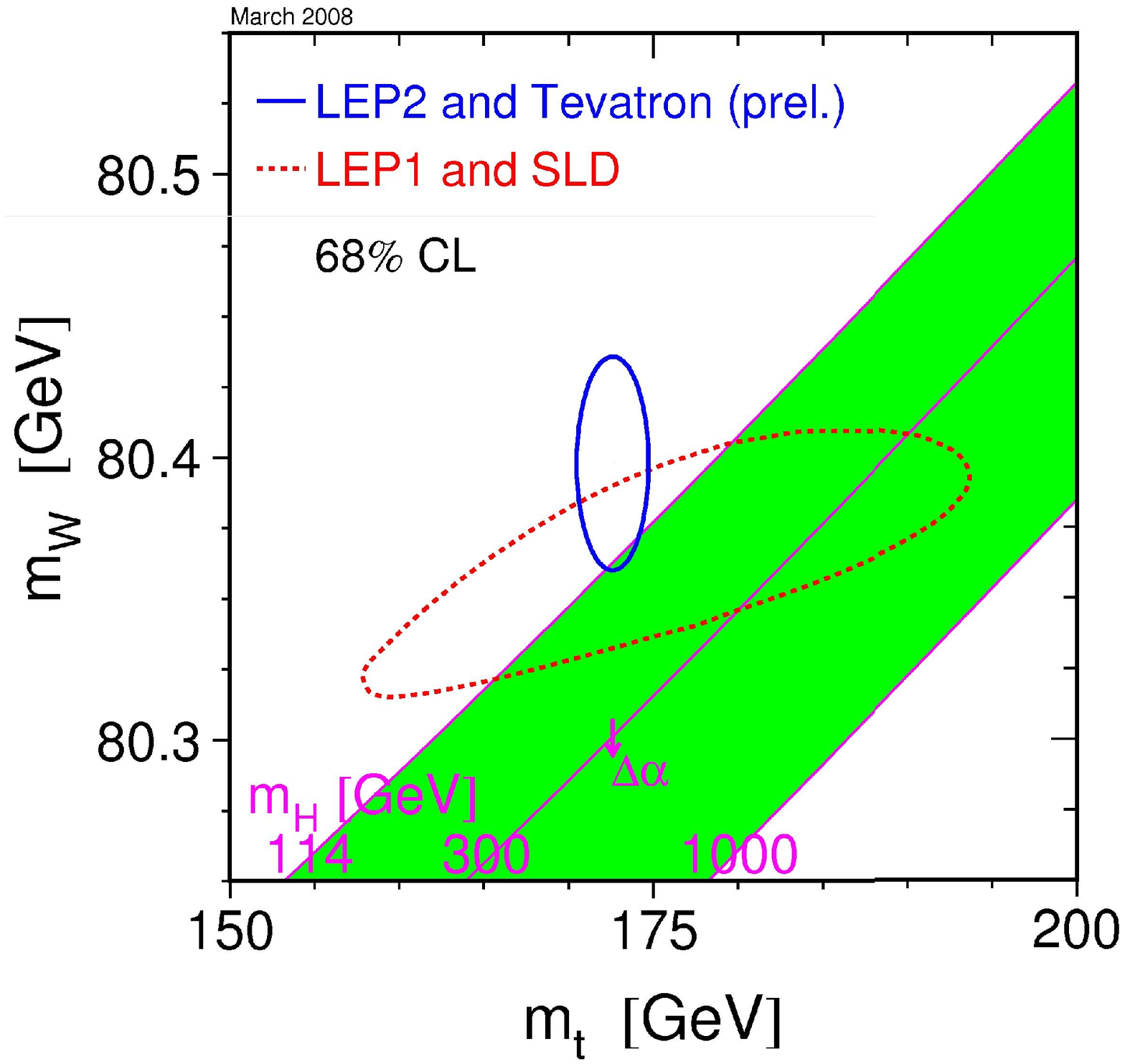}
\caption{$W$ boson mass versus top quark mass plane. The shaded band is consistent with the standard model for 114~GeV$<m_H<$ 1~TeV.}
\label{fig:mw_vs_mtop} 
\end{minipage}
\vspace{-0.1in}
\end{figure}

The top quark is the most massive fundamental particle that we know today. It is the dominant contributor to radiative corrections for many standard model processes. In order to make precise predictions based on the standard model we have to therefore know its mass to high precision. A well-known example for this is the $W$ boson mass. The dominant radiative corrections to the $W$ boson mass come from loops containing top and bottom quarks, which are proportional to the top quark mass squared, and from loops containing Higgs bosons, which are logarithmically dependent on the Higgs boson mass. Thus for every value of the Higgs boson mass $m_H$, the standard model defines a unique line in the plane defined by $W$ boson mass and top quark mass shown in Figure~\ref{fig:mw_vs_mtop}. Vice versa, if we measure the $W$ boson and top quark masses precisely we can constrain the allowed parameter space for the standard model in this plane and therefore limit the allowed range for the Higgs boson mass.

We study top quarks at the Fermilab Tevatron $p\bar{p}$ collider at $\sqrt{s}=1.96$~TeV. The dominant production mechanism for top quarks is top-antitop quark pair productions via quark-antiquark annihilation or gluon fusion. Within the standard model the top quark decays almost exclusively to a $W$ boson and a $b$ quark. The decay modes of the $W$ boson then define the final state signatures. In 6\% of all $t\bar{t}$ decays both $W$ bosons produce electrons or muons in their decay - we call this the dilepton channel. In 38\% of all $t\bar{t}$ decays one $W$ boson produces an electron or a muon in its decay - this is the lepton+jets channel. The remaining 56\% of $t\bar{t}$ decays contain no electrons or muons from $W$ decays - we call this the all-jets channel. In addition to the $W$ bosons, all $t\bar{t}$ decays also produce $b$ and $\bar{b}$ quarks that give rise to jets. Since $b$-flavored hadrons are long-lived, we can identify such jets by the presence of tracks from their secondary decay vertices that do not point back to the primary interaction point. Jets with such tracks or with explicitly reconstructed secondary vertices likely originate from the fragmentation of a $b$ quark and are called $b$-tagged. Requiring at least one $b$-tagged jet in the event is a good way to increase the signal-to-background in the data.

The measurement of the top quark mass relies on the precise reconstruction of the energy of its decay products. We calibrate the measurement of the energies of electron and muons using $Z\to ee$ or $Z\to\mu\mu$ decays. The calibration of the quark energies is a bigger challenge. Quarks manifest themselves as jets in the detector and the first step is to calibrate the measurement of the energy in the jet cone to the energy of all particles from the hadronization of the original quark in the same cone. The response of the calorimeter to the particles is calibrated by balancing the transverse momentum in events with a direct photon and a jet. Then Monte Carlo-derived corrections are applied to take this calibration to the parton level. Dijet events are used to make the calorimeter response to jets uniform in the calorimeter at all momenta. We call this the external jet energy calibration and it is sytematically limited at a precision of about 1.7\%. Alternatively, we can use hadronic decays of $W$ bosons from top decay to calibrate the jet energy scale with top decay events. This determines an additional overall calibration factor for jet energies. We call this the in situ calibration. This calibration is presently statistically limited by the number of hadronic $W$ decays. 

Two basic types of algorithms are currently used for the top mass measurements. The template methods reconstruct one or more observables from each event that serve as estimators for the top quark mass and then compare their distributions to Monte Carlo-derived distributions for a range of top quark masses, the so called templates. More recently, methods have been developed that derive a likelihood as a function of the top quark mass for each event. This has the advantage that every event can be weighted according to its resolution and probability to be from $t\bar{t}$ decays. The most prominant example for the latter is the matrix element method that computes the probability for an event to be produced in $t\bar{t}$ decays as a function of the top quark mass based on inegrating over the tree-level matrix element for $t\bar{t}$ production and decay~\cite{D0_nature}.

It has become standard procedure to calibrate the performance of the measurement algorithms using Monte Carlo generated pseudoexperiments. Events are drawn randomly from simulated signal and background samples according to the number of events observed in the data and the expected or observed signal-to-background. Each pseudoexperiment is analyzed with the same algorithm as the data. The average and root mean square of the measured masses and their pull distribution characterize the performance of the algorithm.

Dilepton decays contain two $b$-jets, two charged leptons and several neutrinos in the final state. This channel has the advantage of a high signal-to-background (3:1) even without using $b$-tagging. Its disadvantage is that the number of measured quantities in these events is insufficient to constrain the kinematics of the events. 

D0 uses two different algorithms to measure the top quark mass in dilepton events based on 1 fb$^{-1}$ of data. The matrix weighting algorithm finds for every top mass the possible top and antitop momenta and then assigns a weight to the events based on parton distribution functions and the lepton energy distribution. The value of the top quark mass at which this weight is maximized is used as the estimator. Its distribution is fit to templates to extract a top quark mass of 175.2$\pm$6.1(stat)$\pm$3.4(syst)~GeV. The neutrino weighting algorithm loops over a range of pseudorapidities for the two neutrinos and uses a $\chi^2$ variable based on the consistency of the sum of the neutrino momenta with the observed missing $p_T$ as weight. A two dimensional template fit to the distributions of mean and rms of the weight curves gives a top quark mass of 172.5$\pm$5.8(stat)$\pm$3.5(syst)~GeV. Combining the two gives 173.7$\pm$5.4(stat)$\pm$3.4(syst)~GeV~\cite{D0_ll}.

CDF use the matrix element method in the dilepton channel. The analysis is carried out on 2 fb$^{-1}$ of data and optimizes the selection with an evolutionary neural network to minimize the statistical uncertainty. The result is 171.2$\pm$2.7(stat)$\pm$2.9(syst)~GeV~\cite{CDF_llME}. CDF also apply the neutrino weighting method. Based on 1.9 fb$^{-1}$ and a two-dimensional template fit to the mass with the largest weight and to the scalar sum of the jet $p_T$s gives 171.6$\pm$3.3(stat)$\pm$3.8(syst)~GeV~\cite{CDF_llvWT}.

In lepton+jets decays we measure everything except the component of the neutrino momentum along the beam direction. With the two mass constraints on the decay products of the $W$ bosons and the requirement for both top quark masses to be equal, a kinematic fit with two constraints is possible. Signal-to-background is typically 1:2 without $b$-tagging and 2:1 after requiring at least one $b$-tagged jet.

D0 use the matrix element method to measure the mass in the lepton+jets channel. D0 integrate over the tree-level matrix elemnts for $t\bar{t}$-production and for $W$+jets production to compute the likelihood that the event originates from either signal or background. The jets from the hadronic $W$ boson decay are used to constrain the jet energy scale calibration. The result $m_t=170.5\pm2.5\mbox{(stat$\oplus$jes)}\pm1.4\mbox{(syst)}$~GeV from the first 0.9~fb$^{-1}$ of data from Run II has already been presented at conferences last year. D0 have analyzed an additional 1.2~fb$^{-1}$ of data resulting in $m_t=173.0\pm1.9\mbox{(stat$\oplus$jes)}\pm1.0\mbox{(syst)}$~GeV. D0 combine the two measurements to obtain $m_t=172.2\pm1.1\mbox{(stat)}\pm1.6\mbox{(syst$\oplus$jes)}$~GeV~\cite{D0_ljME2}. 

CDF also use the matrix element method with in situ calibration of the jet energy scale. CDF only integrate over the $t\bar{t}$-matrix element to compute the likelihood to observe the event from $t\bar{t}$-production. No integration over the background matrix elements is performed. CDF use a neural network discriminant to determine an event-by-event signal-to-background ratio and then subtract the bias in the likelihood expected from the background events in the sample. Based on 1.9~fb$^{-1}$ CDF measure $m_t=172.7\pm1.8\mbox{(stat$\oplus$jes)}\pm1.2\mbox{(syst)}$~GeV~\cite{CDF_ljME}. Alternatively CDF also perform a template fit to the best top quark mass and the best dijet mass from a kinematic fit of the event to the $t\bar{t}$-hypothesis on the same data set. The dijet mass constrains the jet energy scale. With this analysis CDF measure $m_t=171.8\pm1.9\mbox{(stat$\oplus$jes)}\pm1.0\mbox{(syst)}$~GeV~\cite{CDF_llvWT}. 

The all jets channel has the largest branching fraction but also by far the largest background. CDF use events with 6-8 jets, of which at least one has to be $b$-tagged. They fit the best top quark mass and dijet masses from a kinematic fit of the leading six jets to the $t\bar{t}$-hypothesis to templates and obtain $m_t=177.0\pm3.7\mbox{(stat$\oplus$jes)}\pm1.6\mbox{(syst)}$~GeV~\cite{CDF_jj}. 

We use the BLUE method~\cite{BLUE} to combine the best top quark mass measurements from Runs~I and II in each channel by the to experiments, including the correlations between the systematics uncertainties of these measurements. Fig.~\ref{fig:mtop} compares these measurements and Table~\ref{tab:mtop} summarizes the Run~II measurements that enter in this combination and their systematic uncertainties. It can be seen that statistical and systematic uncertainties are of similar magnitude and that the largest source of systematic uncertainty is the jet energy scale calibration. The world average top quark mass is $m_t=172.6\pm1.4$~GeV, which is for the first time below 1\% in precision. The top quark masses determined in the three different final state channels are $169.8\pm3.1$~GeV in the dilepton channel, $172.4\pm1.5$~GeV in the lepton+jets channel, and $177.3\pm3.9$~GeV in the all-jets channel, all consistent with the overall average~\cite{World_average}. 

\begin{table}[t]
\caption{Run II top quark mass measurements and systematic uncertainties used in the world average.\label{tab:mtop}}
\vspace{0.3cm}
\begin{center}
\begin{tabular}{lcccccccc}
\hline
channel    & \multicolumn{2}{c}{dilepton} & \multicolumn{2}{c}{l+jets} & all-jets & $L_{xy}$~\cite{CDF_Lxy} & world \\ 
experiment & CDF & D0 & CDF & D0 & CDF & CDF & average \\ \hline
$\int{\cal L}dt$ & 2.0 & 1.1 & 1.9 & 2.1 & 1.9 & 0.7 & & fb$^{-1}$\\

result     & 171.2 & 173.7 & 172.7 & 172.2 & 177.0 & 180.7 & 172.6 & GeV\\ \hline
jet energy calibration & 1.5 & 3.1 & 1.5 & 1.3 & 2.0 & 0.3 & 0.9 & GeV \\
signal model & 0.7 & 0.8 & 0.6 & 0.7 & 0.6 & 1.4 & 0.5 & GeV \\
background model & 0.4 & 0.6 & 0.6 & 0.4 & 1.0 & 7.2 & 0.4 & GeV \\
fit        & 0.6 & 0.9 & 0.2 & 0.1 & 0.6 & 4.2 & 0.1 & GeV \\
Monte Carlo model & 0.7 & 0.2 & 0.4 & 0.0 & 0.3 & 0.7 & 0.2 & GeV \\ \hline
syst uncertainty & 2.8 & 3.4 & 1.7 & 1.6 & 2.4 & 8.5 & 1.1 & GeV \\
stat uncertainty & 2.7 & 5.4 & 1.2 & 1.1 & 3.3 & 14.5 & 0.8 & GeV \\ \hline
total uncertainty & 3.9 & 6.4 & 2.1 & 1.9 & 4.1 & 16.8 & 1.4 & GeV \\
\hline
\end{tabular}
\end{center}
\vspace{-0.1cm}
\end{table}

The global electroweak fit was repeated with this new world average for the top quark mass as input~\cite{Gruenewald}. The $\chi^2$ is minimized at a value of 17.2 for 13 degrees of freedom, which corresponds to a probability of 19\% for a Higgs boson mass of $87^{+36}_{-27}$~GeV. The 95\% C.L. upper limit for the Higgs boson mass is 160~GeV. Figure~\ref{fig:mw_vs_mtop} shows the measured values of the masses of the top quark and the $W$ boson~\cite{LEPEWWG}. 

 In conclusion, we have measured the top quark mass with 0.8\% precision to be $172.6\pm1.4$~GeV using about 2~fb$^{-1}$ of data from the Fermilab Tevatron collider. By the end of Run~II we expect to accumulate a total of around 6-8 fb$^{-1}$ which will allow us to push the experimental precision of this measurement below 1~GeV.

I would like to thank the organizers for inviting me to present these results at this very exciting and stimulating conference.

\end{document}